\documentclass[preprint]{aastex}
\newcommand{\etal}{et al.}

\newcommand{\cmodel}{{\tt cmodel}}
\newcommand{\Cmodel}{{\tt cmodel}}
\shorttitle{SDSS DR2}
\shortauthors{Abazajian \etal}
\begin{document}
\title{The Second Data Release of the Sloan Digital Sky Survey}
\author{
Kevork Abazajian\altaffilmark{\ref{LANLtheory}},
Jennifer K. Adelman-McCarthy\altaffilmark{\ref{Fermilab}},
Marcel A. Ag\"ueros\altaffilmark{\ref{Washington}},
Sahar S. Allam\altaffilmark{\ref{NMSU}},
Kurt S. J. Anderson\altaffilmark{\ref{APO},\ref{NMSU}},
Scott F. Anderson\altaffilmark{\ref{Washington}},
James Annis\altaffilmark{\ref{Fermilab}},
Neta A. Bahcall\altaffilmark{\ref{Princeton}},
Ivan K. Baldry\altaffilmark{\ref{JHU}},
Steven Bastian\altaffilmark{\ref{Fermilab}},
Andreas Berlind\altaffilmark{\ref{Chicago},\ref{CfCP},\ref{NYU}},
Mariangela Bernardi\altaffilmark{\ref{CMU}},
Michael R. Blanton\altaffilmark{\ref{NYU}},
John J. Bochanski Jr.\altaffilmark{\ref{Washington}},
William N. Boroski\altaffilmark{\ref{Fermilab}},
John W. Briggs\altaffilmark{\ref{Chicago}},
J. Brinkmann\altaffilmark{\ref{APO}},
Robert J. Brunner\altaffilmark{\ref{Illinois}},
Tam\'as Budav\'ari\altaffilmark{\ref{JHU}},
Larry N. Carey\altaffilmark{\ref{Washington}},
Samuel Carliles\altaffilmark{\ref{JHU}},
Francisco J. Castander\altaffilmark{\ref{Barcelona}},
A. J. Connolly\altaffilmark{\ref{Pitt}},
Istv\'an Csabai\altaffilmark{\ref{Eotvos},\ref{JHU}},
Mamoru Doi\altaffilmark{\ref{IoAUT}},
Feng Dong\altaffilmark{\ref{Princeton}},
Daniel J. Eisenstein\altaffilmark{\ref{Arizona}},
Michael L. Evans\altaffilmark{\ref{Washington}},
Xiaohui Fan\altaffilmark{\ref{Arizona}},
Douglas P. Finkbeiner\altaffilmark{\ref{Princeton}},
Scott D. Friedman\altaffilmark{\ref{STScI}},
Joshua A. Frieman\altaffilmark{\ref{Fermilab},\ref{Chicago},\ref{CfCP}},
Masataka Fukugita\altaffilmark{\ref{ICRRUT}},
Roy R. Gal\altaffilmark{\ref{JHU}},
Bruce Gillespie\altaffilmark{\ref{APO}},
Karl Glazebrook\altaffilmark{\ref{JHU}},
Jim Gray\altaffilmark{\ref{Microsoft}},
Eva K. Grebel\altaffilmark{\ref{MPIA}},
James E. Gunn\altaffilmark{\ref{Princeton}},
Vijay K. Gurbani\altaffilmark{\ref{Fermilab},\ref{Lucent2}},
Patrick B. Hall\altaffilmark{\ref{Princeton}},
Masaru Hamabe\altaffilmark{\ref{JapanWomen}},
Frederick H. Harris\altaffilmark{\ref{NOFS}},
Hugh C. Harris\altaffilmark{\ref{NOFS}},
Michael Harvanek\altaffilmark{\ref{APO}},
Timothy M. Heckman\altaffilmark{\ref{JHU}},
John S. Hendry\altaffilmark{\ref{Fermilab}},
Gregory S. Hennessy\altaffilmark{\ref{USNO}},
Robert B. Hindsley\altaffilmark{\ref{NRL}},
Craig J. Hogan\altaffilmark{\ref{Washington}},
David W. Hogg\altaffilmark{\ref{NYU}},
Donald J. Holmgren\altaffilmark{\ref{Fermilab}},
Shin-ichi Ichikawa\altaffilmark{\ref{NAOJ}},
Takashi Ichikawa\altaffilmark{\ref{TohokuU}},
\v{Z}eljko Ivezi\'{c}\altaffilmark{\ref{Princeton}},
Sebastian Jester\altaffilmark{\ref{Fermilab}},
David E. Johnston\altaffilmark{\ref{Chicago},\ref{CfCP},\ref{Princeton}},
Anders M. Jorgensen\altaffilmark{\ref{LANL}},
Stephen M. Kent\altaffilmark{\ref{Fermilab}},
S. J. Kleinman\altaffilmark{\ref{APO}},
G. R. Knapp\altaffilmark{\ref{Princeton}},
Alexei Yu. Kniazev\altaffilmark{\ref{MPIA}},
Richard G. Kron\altaffilmark{\ref{Chicago},\ref{Fermilab}},
Jurek Krzesinski\altaffilmark{\ref{APO},\ref{MSO}},
Peter Z. Kunszt\altaffilmark{\ref{JHU},\ref{CERN}},
Nickolai Kuropatkin\altaffilmark{\ref{Fermilab}},
Donald Q. Lamb\altaffilmark{\ref{Chicago},\ref{EFI}},
Hubert Lampeitl\altaffilmark{\ref{Fermilab}},
Brian C. Lee\altaffilmark{\ref{LBL}},
R. French Leger\altaffilmark{\ref{Fermilab}},
Nolan Li\altaffilmark{\ref{JHU}},
Huan Lin\altaffilmark{\ref{Fermilab}},
Yeong-Shang Loh\altaffilmark{\ref{Princeton},\ref{Colorado}},
Daniel C. Long\altaffilmark{\ref{APO}},
Jon Loveday\altaffilmark{\ref{Sussex}},
Robert H. Lupton\altaffilmark{\ref{Princeton}},
Tanu Malik\altaffilmark{\ref{JHU}},
Bruce Margon\altaffilmark{\ref{STScI}},
Takahiko Matsubara\altaffilmark{\ref{Nagoya}},
Peregrine M. McGehee\altaffilmark{\ref{NMSU},\ref{LANL2}},
Timothy A. McKay\altaffilmark{\ref{Michigan}},
Avery Meiksin\altaffilmark{\ref{Edinburgh}},
Jeffrey A. Munn\altaffilmark{\ref{NOFS}},
Reiko Nakajima\altaffilmark{\ref{Penn}},
Thomas Nash\altaffilmark{\ref{Fermilab}},
Eric H. Neilsen, Jr.\altaffilmark{\ref{Fermilab}},
Heidi Jo Newberg\altaffilmark{\ref{RPI}},
Peter R. Newman\altaffilmark{\ref{APO}},
Robert C. Nichol\altaffilmark{\ref{CMU}},
Tom Nicinski\altaffilmark{\ref{Fermilab},\ref{CMCElectronics}},
Maria Nieto-Santisteban\altaffilmark{\ref{JHU}},
Atsuko Nitta\altaffilmark{\ref{APO}},
Sadanori Okamura\altaffilmark{\ref{DoAUT}},
William O'Mullane\altaffilmark{\ref{JHU}},
Jeremiah P. Ostriker\altaffilmark{\ref{Princeton}},
Russell Owen\altaffilmark{\ref{Washington}},
Nikhil Padmanabhan\altaffilmark{\ref{Princetonphys}},
John Peoples\altaffilmark{\ref{Fermilab}},
Jeffrey R. Pier\altaffilmark{\ref{NOFS}},
Adrian C. Pope\altaffilmark{\ref{JHU}},
Thomas R. Quinn\altaffilmark{\ref{Washington}},
Gordon T. Richards\altaffilmark{\ref{Princeton}},
Michael W. Richmond\altaffilmark{\ref{RIT}},
Hans-Walter Rix\altaffilmark{\ref{MPIA}},
Constance M. Rockosi\altaffilmark{\ref{Washington}},
David J. Schlegel\altaffilmark{\ref{Princeton}},
Donald P. Schneider\altaffilmark{\ref{PSU}},
Ryan Scranton\altaffilmark{\ref{Pitt}},
Maki Sekiguchi\altaffilmark{\ref{JPG}},
Uros Seljak\altaffilmark{\ref{Princeton}},
Gary Sergey\altaffilmark{\ref{Fermilab}},
Branimir Sesar\altaffilmark{\ref{Zagreb}},
Erin Sheldon\altaffilmark{\ref{Chicago},\ref{CfCP}},
Kazu Shimasaku\altaffilmark{\ref{DoAUT}},
Walter A. Siegmund\altaffilmark{\ref{Hawaii}},
Nicole M. Silvestri\altaffilmark{\ref{Washington}},
J. Allyn Smith\altaffilmark{\ref{Wyoming},\ref{LANL}},
Vernesa Smol\v{c}i\'{c}\altaffilmark{\ref{Zagreb}},
Stephanie A. Snedden\altaffilmark{\ref{APO}},
Albert Stebbins\altaffilmark{\ref{Fermilab}},
Chris Stoughton\altaffilmark{\ref{Fermilab}},
Michael A. Strauss\altaffilmark{\ref{Princeton}},
Mark SubbaRao\altaffilmark{\ref{Chicago},\ref{Adler}},
Alexander S. Szalay\altaffilmark{\ref{JHU}},
Istv\'an Szapudi\altaffilmark{\ref{Hawaii}},
Paula Szkody\altaffilmark{\ref{Washington}},
Gyula P. Szokoly\altaffilmark{\ref{MPIEP}},
Max Tegmark\altaffilmark{\ref{Penn}},
Luis Teodoro\altaffilmark{\ref{LANLtheory}},
Aniruddha R. Thakar\altaffilmark{\ref{JHU}},
Christy Tremonti\altaffilmark{\ref{Arizona}},
Douglas L. Tucker\altaffilmark{\ref{Fermilab}},
Alan Uomoto\altaffilmark{\ref{JHU},\ref{CarnegieObs}},
Daniel E. Vanden Berk\altaffilmark{\ref{Pitt}},
Jan Vandenberg\altaffilmark{\ref{JHU}},
Michael S. Vogeley\altaffilmark{\ref{Drexel}},
Wolfgang Voges\altaffilmark{\ref{MPIEP}},
Nicole P. Vogt\altaffilmark{\ref{NMSU}},
Lucianne M. Walkowicz\altaffilmark{\ref{Washington}},
Shu-i Wang\altaffilmark{\ref{Chicago}},
David H. Weinberg\altaffilmark{\ref{OSU}},
Andrew A. West\altaffilmark{\ref{Washington}},
Simon D.M. White\altaffilmark{\ref{MPA}},
Brian C. Wilhite\altaffilmark{\ref{Chicago}},
Yongzhong Xu\altaffilmark{\ref{LANLtheory}},
Brian Yanny\altaffilmark{\ref{Fermilab}},
Naoki Yasuda\altaffilmark{\ref{ICRRUT}},
Ching-Wa Yip\altaffilmark{\ref{Pitt}},
D. R. Yocum\altaffilmark{\ref{Fermilab}},
Donald G. York\altaffilmark{\ref{Chicago},\ref{EFI}},
Idit Zehavi\altaffilmark{\ref{Arizona}},
Stefano Zibetti\altaffilmark{\ref{MPA}},
Daniel B. Zucker\altaffilmark{\ref{MPIA}}
}

\altaffiltext{1}{
Theoretical Division, MS B285, Los Alamos National Laboratory, Los Alamos, NM 87545
\label{LANLtheory}}

\altaffiltext{2}{
Fermi National Accelerator Laboratory, P.O. Box 500, Batavia, IL 60510
\label{Fermilab}}

\altaffiltext{3}{
Department of Astronomy, University of Washington, Box 351580, Seattle, WA
98195
\label{Washington}}

\altaffiltext{4}{
New Mexico State University, Department of Astronomy, P.O. Box 30001, Dept
4500, Las Cruces, NM 88003
\label{NMSU}}

\altaffiltext{5}{
Apache Point Observatory, P.O. Box 59, Sunspot, NM 88349
\label{APO}}

\altaffiltext{6}{
Department of Astrophysical Sciences, Princeton University, Princeton, NJ
08544
\label{Princeton}}

\altaffiltext{7}{
Center for Astrophysical Sciences, Department of Physics \& Astronomy, Johns
Hopkins University, Baltimore, MD 21218
\label{JHU}}

\altaffiltext{8}{
Department of Astronomy and Astrophysics, The University of Chicago, 5640 S.
Ellis Ave., Chicago, IL 60637
\label{Chicago}}

\altaffiltext{9}{
Center for Cosmological Physics, The University of Chicago,
5640 South Ellis Ave., Chicago, IL 60637
\label{CfCP}}

\altaffiltext{10}{
Center for Cosmology and Particle Physics,
Department of Physics,
New York University,
4 Washington Place,
New York, NY 10003
\label{NYU}}

\altaffiltext{11}{
Department of Physics, Carnegie Mellon University, Pittsburgh, PA 15213
\label{CMU}}

\altaffiltext{12}{
Department of Astronomy
University of Illinois
1002 W. Green Street, Urbana, IL 61801
\label{Illinois}}

\altaffiltext{13}{Institut d'Estudis Espacials de Catalunya/CSIC, Gran Capita 2-4,
08034 Barcelona, Spain
\label{Barcelona}}

\altaffiltext{14}{
Department of Physics and Astronomy, University of Pittsburgh, 3941 O'Hara
St., Pittsburgh, PA 15260
\label{Pitt}}

\altaffiltext{15}{
Department of Physics, E\"{o}tv\"{o}s University, Budapest, Pf.\ 32,
Hungary, H-1518
\label{Eotvos}}

\altaffiltext{16}{Institute of Astronomy, School
of Science, University of Tokyo,
 2-21-1 Osawa, Mitaka, Tokyo 181-0015, Japan
\label{IoAUT}}

\altaffiltext{17}{
Steward Observatory, 933 N. Cherry Ave, Tucson, AZ 85721
\label{Arizona}}

\altaffiltext{18}{
Space Telescope Science Institute, 3700 San Martin Drive, Baltimore, MD
21218
\label{STScI}}

\altaffiltext{19}{Institute for Cosmic Ray Research, University of Tokyo, 5-1-5 Kashiwa-no-Ha,
 Kashiwa City, Chiba 277-8582, Japan
\label{ICRRUT}}

\altaffiltext{20}{
Microsoft Research, 455 Market Street, Suite 1690, San Francisco, CA 94105
\label{Microsoft}}

\altaffiltext{21}{
Max-Planck Institute for Astronomy, K\"onigstuhl 17, D-69117 Heidelberg,
Germany
\label{MPIA}}

\altaffiltext{22}{
Lucent Technologies, 2000 Lucent Lane, Naperville, IL 60566
\label{Lucent2}}

\altaffiltext{23}{
Dept. Mathematical \& Physical Sciences,
Japan Women's University
2-8-1, Mejirodai, Bunkyo-ku, Tokyo 112-8681, Japan
\label{JapanWomen}}

\altaffiltext{24}{
U.S. Naval Observatory, Flagstaff Station, P.O. Box 1149, Flagstaff, AZ
86002-1149
\label{NOFS}}

\altaffiltext{25}{
US Naval Observatory, 3540 Mass Ave NW, Washington, DC 20392
\label{USNO}}

\altaffiltext{26}{National Astronomical Observatory, 2-21-1 Osawa, Mitaka, Tokyo 181-8588,
Japan
\label{NAOJ}}

\altaffiltext{27}{Astronomical Institute, Tohoku University, Aramaki, Aoba, Sendai 980-8578,
Japan
\label{TohokuU}}

\altaffiltext{28}{
ISR-4, MS D448, Los Alamos National Laboratory, Los Alamos, NM 87545
\label{LANL}}

\altaffiltext{29}{
Mt. Suhora Observatory, Cracow Pedagogical University, ul. Podchorazych 2,
30-084 Cracow, Poland
\label{MSO}}

\altaffiltext{30}{
CERN, IT Division, 1211 Geneva 23, Switzerland
\label{CERN}}

\altaffiltext{31}{
Enrico Fermi Institute, The University of Chicago, 5640 S. Ellis Ave.,
Chicago, IL 60637
\label{EFI}}

\altaffiltext{32}{
Lawrence Berkeley National Laboratory, One Cyclotron Rd.,
Berkeley CA 94720-8160
\label{LBL}}

\altaffiltext{33}{
Center for Astrophysics and Space Astronomy, University of Colorado,
Boulder, CO 80309
\label{Colorado}}

\altaffiltext{34}{
Astronomy Centre, University of Sussex, Falmer, Brighton BN1 9QJ, United
Kingdom
\label{Sussex}}

\altaffiltext{35}{
Department of Physics and Astrophysics
 Nagoya University
 Chikusa, Nagoya 464-8602,
 Japan
\label{Nagoya}}

\altaffiltext{36}{
SNS-4, MS H820, Los Alamos National Laboratory, Los Alamos, NM 87545
\label{LANL2}}

\altaffiltext{37}{
Department of Physics, University of Michigan, 500 East University Ave., Ann
Arbor, MI 48109
\label{Michigan}}

\altaffiltext{38}{
Institute for Astronomy
Royal Observatory
Blackford Hill
Edinburgh EH9 3HJ
Scotland
\label{Edinburgh}}

\altaffiltext{39}{
Department of Physics, University of Pennsylvania, Philadelphia, PA 19104
\label{Penn}}

\altaffiltext{40}{
Department of Physics, Applied Physics, and Astronomy, Rensselaer
Polytechnic Institute, Troy, NY 12180
\label{RPI}}

\altaffiltext{41}{
    CMC Electronics Aurora
    43W752 Route 30
    Sugar Grove, IL 60554
\label{CMCElectronics}}

\altaffiltext{42}{Department of Astronomy and Research Center for the Early Universe, School
of Science, University of Tokyo,
 7-3-1 Hongo, Bunkyo, Tokyo 113-0033, Japan
\label{DoAUT}}

\altaffiltext{43}{
Joseph Henry Laboratories, Princeton University, Princeton, NJ
08544
\label{Princetonphys}}

\altaffiltext{44}{
Physics Department, Rochester Institute of Technology, 84 Lomb Memorial
Drive, Rochester, NY 14623-5603
\label{RIT}}

\altaffiltext{45}{
Department of Astronomy and Astrophysics, the Pennsylvania State
University, University Park, PA 16802
\label{PSU}}

\altaffiltext{46}{
Japan Participation Group, 
c/o Institute for Cosmic Ray Research, University of Tokyo, 5-1-5
Kashiwa-no-Ha, Kashiwa City, Chiba 277-8582, Japan
\label{JPG}}

\altaffiltext{47}{University of Zagreb, 
Department of Physics, Bijeni\v{c}ka cesta 32, 
10000 Zagreb, Croatia
\label{Zagreb}}

\altaffiltext{48}{
Institute for Astronomy, 2680 Woodlawn Road, Honolulu, HI 96822
\label{Hawaii}}

\altaffiltext{49}{
University of Wyoming, Dept. of Physics \& Astronomy, Laramie, WY 82071
\label{Wyoming}}

\altaffiltext{50}{
Adler Planetarium and Astronomy Museum
1300 Lake Shore Drive
Chicago, IL 60605
\label{Adler}}

\altaffiltext{51}{
Max-Planck-Institut f\"ur extraterrestrische Physik, 
Giessenbachstrasse 1, D-85741 Garching, Germany
\label{MPIEP}}

\altaffiltext{52}{
Carnegie Observatories
813 Santa Barbara St.
Pasadena, CA  91101
\label{CarnegieObs}}

\altaffiltext{53}{
Department of Physics, Drexel University, Philadelphia, PA 19104
\label{Drexel}}

\altaffiltext{54}{
Department of Astronomy, Ohio State University, Columbus, OH 43210
\label{OSU}}

\altaffiltext{55}{
Max Planck Institute for Astrophysics, Karl Schwarzschildstrasse 1,
D-85748 Garching, Germany
\label{MPA}}

\altaffiltext{56}{
Code 7215, Remote Sensing Division
Naval Research Laboratory
4555 Overlook Ave SW
Washington, DC 20392
\label{NRL}}
\begin{abstract}
The Sloan Digital Sky Survey has validated and made publicly available
its Second Data Release.  This data release consists of 3324 square degrees of
five-band ($u\,g\,r\,i\,z$) imaging data with photometry for over 88
million unique objects, 367,360 spectra 
of galaxies, 
quasars, stars and calibrating blank sky patches selected over
2627
degrees of this area, and tables of measured parameters from these
data.  
The imaging data reach a depth of $r \approx 22.2$ (95\% completeness
limit for point sources) and are photometrically and astrometrically
calibrated to 2\% rms and 100 milli-arcsec rms per coordinate,
respectively.  The imaging data have all been processed through a new
version of the SDSS imaging pipeline, in which the most important
improvement since the last data release is fixing an error in the
model fits to each object.  The 
result is that model magnitudes are now a good proxy for point spread
function (PSF) magnitudes for point sources, and Petrosian magnitudes
for extended sources.
The spectroscopy extends from 3800\AA\ to 9200\AA\ at a resolution of
2000.  The spectroscopic software now repairs a systematic error in
the radial velocities of certain types of stars, and has substantially
improved spectrophotometry.  All data included in the SDSS Early Data
Release and First Data Release are reprocessed with the improved
pipelines, and included in the Second Data Release.  Further
characteristics of the data are 
described, as are the data products themselves and the tools for
accessing them. 
\end{abstract}
\keywords{Atlases---Catalogs---Surveys}

\section{The Sloan Digital Sky Survey}

The Sloan Digital Sky Survey (SDSS; York \etal\ 2000) is an imaging and spectroscopic survey of
the high Galactic latitude sky visible from the Northern hemisphere.
The principal survey goals are to measure the large-scale distribution
of galaxies and quasars and to produce an imaging and spectroscopic
legacy for the astronomical community.  The SDSS data have been used
in well over 200 
refereed papers to date on subjects ranging from the colors of
asteroids (Ivezi\'c \etal\ 2002) to magnetic white dwarfs (Schmidt
\etal\ 2003) to structures in the Galactic halo (Newberg \etal\ 2003)
to the star-formation history of galaxies (Kauffmann \etal\ 2003) to Type
II quasars (Zakamska \etal\ 2003) to the large-scale distribution of
galaxies (Pope \etal\ 2004; Tegmark \etal\ 2004).  The survey uses a dedicated 2.5m
telescope with a three-degree field of view at Apache Point
Observatory, New Mexico.  A 120 mega-pixel camera (Gunn \etal\ 1998)
images in five broad bands ($u,g,r,i$ and $z$; Fukugita \etal\
1996; Stoughton \etal\ 2002) on clear moonless nights of good seeing.
These data are photometrically calibrated using an
auxiliary 20-inch telescope with a $40^\prime \times 40^\prime$
imager, which determines 
the photometricity of each night (Hogg \etal\ 2001), and measures the
extinction and photometric zeropoint using a network of 
standard stars (Smith \etal\ 2002).  

The imaging data are processed
through a series of pipelines that locate and measure the properties of
all detected 
objects (Lupton \etal\ 2001) and carry out photometric
and astrometric calibration (Pier \etal\ 2003).  From the resulting
catalogs of objects, complete catalogs of galaxies (Eisenstein \etal\
2001; Strauss \etal\ 2002) and quasar candidates (Richards \etal\
2002) are selected for spectroscopic followup, and are assigned to
spectroscopic tiles of diameter 3 degrees (Blanton \etal\
2003).  Spectroscopy is performed on nights with moonlight, mild cloud
cover, and/or poor seeing using a pair of double spectrographs with
coverage from 3800--9200\AA, and resolution $\lambda / \Delta \lambda$
of roughly 2000.  A plug plate for each tile holds 640 optical fibers
of $3''$ entrance aperture which feed the spectrographs, together
with eleven coherent fiber bundles to image guide stars.  Because
of the diameter of the cladding holding the optical fibers,
spectroscopy cannot be carried out for objects separated by less
than $55''$ on a given plate. 

\section{The Second Data Release}

  A high-level overview of the SDSS may be found in York \etal\ (2000), while
many of the details of the software and data products may be found in
Stoughton \etal\ (2002; hereafter the EDR paper).  The latter paper
also describes our Early 
Data Release (EDR), which consisted mostly of data taken during our
commissioning period.  The First Data Release was made public in April
2003; Abazajian \etal\ (2003; hereafter the DR1 paper) describe these data and
give further details and updates on the quality of the data and its
processing.  The current paper describes the Second Data Release of
the SDSS (DR2), which was made available to the public on 15 March
2004.   The properties of DR2 are summarized in
Table~\ref{tab:characteristics}.  The DR2 
footprint is defined by all non-repeating survey-quality imaging runs within the a priori
defined elliptical survey area in the Nothern Galactic Cap, and three
stripes in the Southern Galactic Cap (York \etal\ 2000) obtained prior to 1
July 2002, and the spectroscopy associated with that area obtained
before that date.  In fact, 34 square degrees of DR2 imaging data in
the Northern Galactic Cap lie outside this
ellipse.  While the DR2 scans do not repeat a given area of sky,
they do overlap to some extent, and the data in the overlaps are
included in DR2 as well.  The DR2 includes reprocessing of
all data included in DR1, and those data in EDR that
pass our data-quality criteria for the official survey. 

\begin{figure}[t]\centering
\includegraphics[width=12cm]{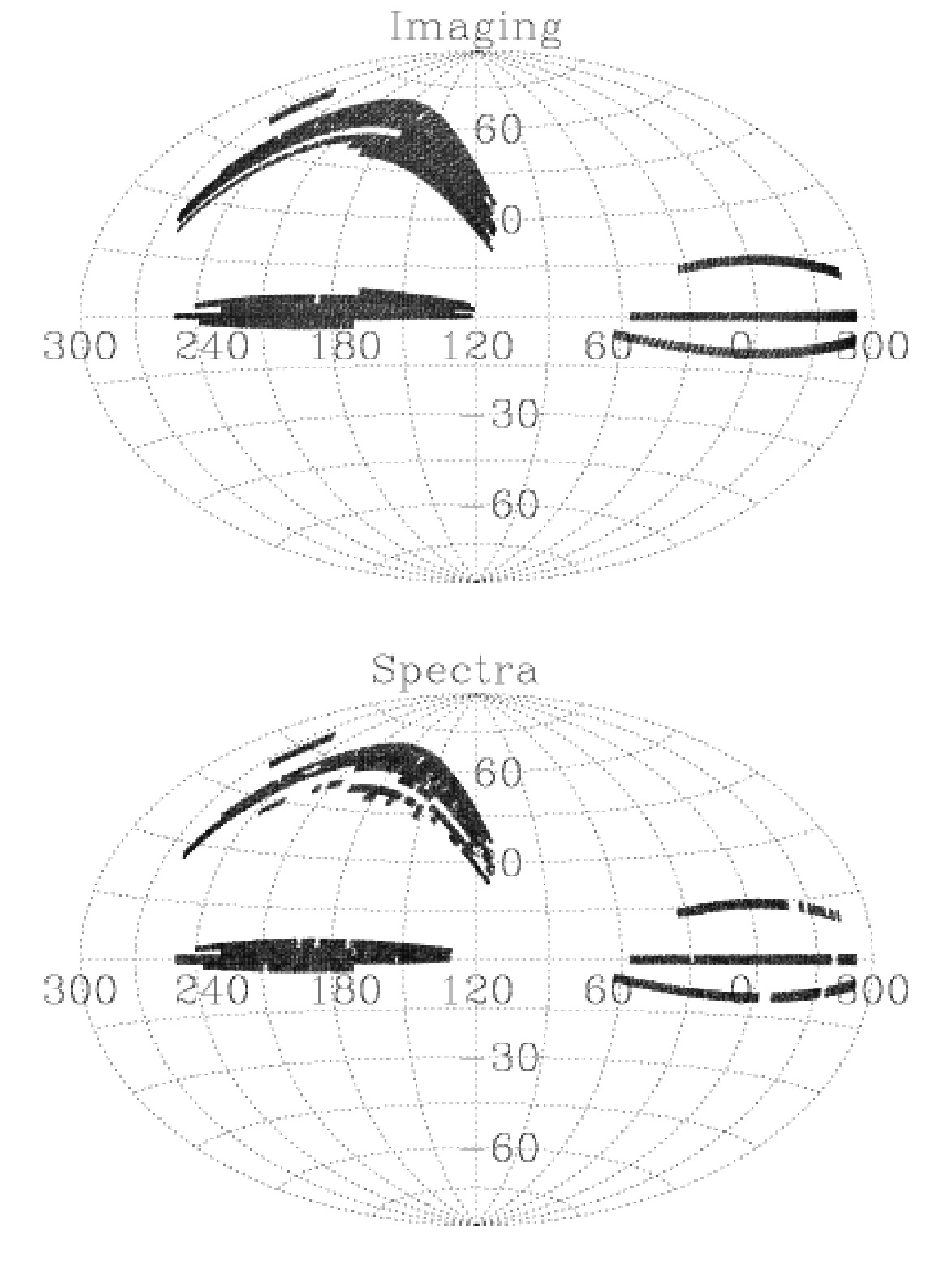}
\caption{The
distribution on the sky of the imaging scans and spectroscopic plates
included in DR2.  This is an Aitoff projection in equatorial
coordinates.  The total sky area covered by the imaging is
3324 square degrees, and by the spectroscopy is 2627 square degrees. 
\label{fig:dr1image}}\end{figure}

The sky coverage of the imaging and spectroscopic data
that make up DR2 is given in Figure~\ref{fig:dr1image}.  The effective
areas of the two are 3324 and 2627 square degrees, respectively.   The
natural unit of imaging data is a {\em run}; the DR2 contains data
from 105 runs in the {\tt best} database, and 105 runs in the {\tt
  target} database ({\tt best} and {\tt target} are defined in
\S~\ref{sec:imaging}).  Similarly, the natural unit of spectroscopic data
is a {\em plate} of 640 spectra each (of which 32 are devoted to
background sky determination); the DR2 contains data from
574 plates. 

The DR2 data are available via links off the web site {\tt
  http://www.sdss.org/dr2}, which also includes extensive technical
information about the SDSS data, and should be checked for errata and
caveats to the data.  The types of data that are available
are described in detail in the DR1 and EDR papers, and on the above
web site.  There are two principal ways to access the data.  The
first, the Data Archive Server (DAS), allows one to download the FITS
files containing the imaging and spectroscopic catalogs, the images
themselves, and the spectra.  This is the appropriate place to go to
download large quantities of data in bulk. 
The second option, the Catalog Archive Server
(CAS), allows one to perform database queries by object attributes
and to obtain finding charts of given regions of sky and plots of the
spectra.  The CAS also provides pointers to the survey images and
spectra in FITS format.  

The EDR paper (Stoughton \etal\ 2002) describes the SDSS data in
detail.  We do not repeat that description here, but put emphasis on
changes since DR1 and new-found problems in imaging
(\S~\ref{sec:imaging}), spectroscopy (\S~\ref{sec:spectroscopy}), and
target selection (\S~\ref{sec:target}).  An Appendix describes the
conversion between magnitudes, fluxes, and counts in the imaging
data. 

\begin{deluxetable}{lr}
\tablecaption{Characteristics of the SDSS Second Data
  Release (DR2)\label{tab:characteristics}}

\startdata

\cutinhead{\bf Imaging} 

 Footprint area & 3324\ deg$^2$ \\
 Imaging catalog & 88 million unique objects \\
 Magnitude limits:\tablenotemark{a} \\
\qquad $u$      & 22.0 \\
\qquad $g$      & 22.2 \\
\qquad $r$      & 22.2 \\
\qquad $i$      & 21.3 \\
\qquad $z$      & 20.5 \\
Median PSF width     & $1.4^{\prime\prime}$ in $r$ \\
 RMS photometric calibration errors: \\
\qquad   $r$ & 2\% \\
\qquad   $u-g$ & 3\% \\
\qquad   $g-r$ & 2\% \\
\qquad   $r-i$ & 2\% \\
\qquad   $i-z$ & 3\% \\
 Astrometry    & $< 0.1^{\prime\prime}$ rms absolute per coordinate \\

\cutinhead{\bf Spectroscopy}\\

Footprint area  & 2627\ deg$^2$\\
Wavelength Coverage & 3800--9200\AA\\
 Resolution     &1800--2100 \\
 Signal-to-noise ratio &$>4$ per pixel at $g=20.2$\\
 Wavelength calibration & $<5$ km sec$^{-1}$\\
 Redshift accuracy & 30 km sec$^{-1}$ rms for Main galaxies\\
 Number of spectra& 367,360\\
\enddata

\tablenotetext{a}{95\% completeness for point sources in
  typical seeing; 50\% completeness numbers are typically 0.4 mag
fainter (DR1 paper).}
\end{deluxetable}

\section{The SDSS Imaging Data}
\label{sec:imaging}

The SDSS imaging pipelines have evolved as the survey has progressed,
leading to continual
improvements in the measured quantities.  However, this means that
much of the spectroscopic targetting is based on 
reductions carried out with  old
versions of the imaging pipelines.  For this reason, we release two
versions of the imaging data for each area of the sky.  The {\tt
target} version is that which was used for defining spectroscopic
targets of a given region, while the {\tt best} version uses the latest
version of the imaging pipelines.  In some cases, improved data (e.g.,
with better seeing) were taken in a given region of sky after
targetting was done, in which case the {\tt target} and {\tt best}
data are independent runs.   A total of 183 square degrees of sky are
different runs between {\tt target} and {\tt best}, the majority along
the Equatorial Stripe in the Fall sky. 

  The quality of the imaging data is described in the EDR and DR1
papers; the distribution of seeing, sky brightness, and depth for the
DR2 data are no different than for DR1, nor are the accuracies of the
astrometric and photometric calibrations.  The {\tt best} reductions do,
however, have some substantial improvements over what was included in
DR1.  
We now describe these improvements, and give further caveats of
problems that have since come to light.  

\subsection{Model Magnitudes}
\label{sec:modelmag}

The computation of model magnitudes in 
the DR1 and EDR processing had a serious bug.  The image of each
object detected in any 
of the five bands in the imaging data is fit to de Vaucouleurs
($I(\theta) \propto \exp[-(\theta/\theta_0)^{1/4}]$) and exponential
($I(\theta) \propto \exp[-\theta/\theta_0]$) radial profiles of arbitrary axis
ratio and inclination, convolved with the local Point Spread Function
(PSF).  However, these fits used an incorrect model of the PSF,
which caused systematic errors in the fit parameters, 
especially for objects of small scale size $\theta_0$ (i.e., close to
a PSF).  For each of these model fits, the code determined an aperture
correction to force the exponential and de Vaucouleurs magnitudes to
equal the PSF magnitude for stars; this correction was then applied to
{\em all} objects.  Because 
of this software error, this aperture correction was large, 0.2 magnitudes.  Thus
the model magnitudes of galaxies were systematically too large by
typically 0.2 magnitudes.  In the mean,
exponential scale lengths {\tt rexp} were overestimated by $\sim
0.1''-0.2''$ for objects with ${\tt rexp} < 6''$, while de Vaucouleurs
scale lengths {\tt rdeV} were overestimated by a factor 1.25 for large
objects, and as much as a factor of 2 for objects with ${\tt rdeV} <
0.5''$.  

  This error has been fixed in the latest version of the pipeline,
and has been extensively tested with simulations.  The code now also
takes the best fit exponential and de Vaucouleurs 
  fits in each band and asks for the linear combination of
  the two that best fits the image.  The coefficient (clipped between
  zero and one) of the de Vaucouleurs term is stored in the quantity
  {\tt fracDeV}\footnote{Due to an accident of the history this parameter
    is misleadingly termed {\tt fracPSF} in the flat files of the DAS.}.  
This allows us to define a {\em composite flux}: 
\begin{equation} 
F_{composite} = {\tt fracDeV} \times {F_{deV}} + (1 - {\tt fracDeV}) \times {
  F_{exp}},
\end{equation}
where $F_{deV}$ and $F_{exp}$ are the de Vaucouleurs and exponential
fluxes ({\em not} magnitudes) of the object in question.  The
magnitude derived from $F_{composite}$ is referred to below as the
{\tt \cmodel} magnitude (as distinct from the {\tt model} magnitude, based on
the better-fitting of the exponential and de Vaucouleurs models in the
$r$ band; see the EDR paper). 

\begin{figure}[t]\centering
\includegraphics[width=12cm,angle=270]{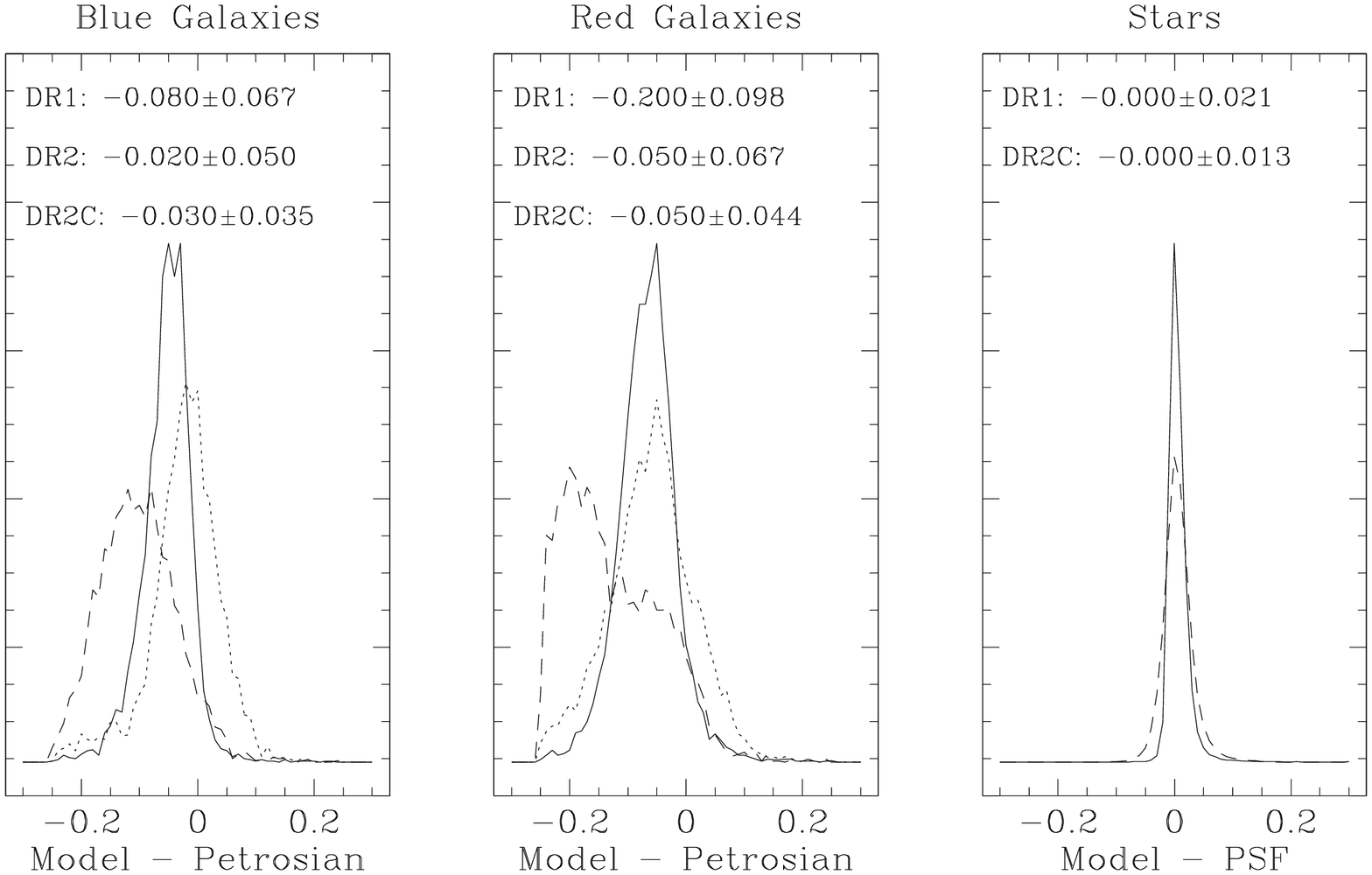}
\caption{Improvements in model magnitudes for stars and galaxies in
  the DR2 reductions.  The first panel shows the distribution of
  differences between $r$ band model and Petrosian magnitudes for red ($u-r >
  2.22$; Strateva \etal\ 2001) galaxies
  brighter than $r_{Petro} = 19$; the three curves are for the old
  (DR1) reductions (dashed), the current reductions using model
  magnitudes (dotted; DR2), and the current reductions using {\tt cmodel}
  magnitudes (solid; DR2C).  The mode and standard
  deviation (based on the interquartile range) of each distribution
  are given.  
  The bias in model magnitudes in the DR1 reductions is apparent.  The
  second panel shows the same quantities for blue ($u-r<2.22$)
  galaxies.  The third panel shows the difference between
  \cmodel\ and PSF magnitudes for $r_{PSF} < 20$ stars, in the DR1
  (dotted) and DR2 (solid) reductions; the width of the distribution has
  decreased by 40\% with the new reductions. 
\label{fig:modelmag}}\end{figure}

  With these changes in place, there is now excellent agreement between
\cmodel\ and Petrosian magnitudes of galaxies, and \cmodel\ and PSF
magnitudes of stars (Figure~\ref{fig:modelmag}). \Cmodel\ and
Petrosian magnitudes are not expected 
to be identical, of course; as Strauss \etal\ (2002) describe, the
Petrosian aperture excludes the outer parts of galaxy profiles,
especially for elliptical galaxies.  As a consequence, there is an
offset of 0.05--0.1 mag between \cmodel\ and Petrosian
magnitudes of bright galaxies, depending on the photometric bandpass
and the type of galaxy.  The rms scatter between model and Petrosian
magnitudes at the bright end is now between 0.05 and 0.08 magnitudes,
depending on bandpass; the scatter between \cmodel\ and Petrosian
magnitudes for all galaxies is smaller, 0.03 to 0.05 magnitudes.  For
comparison, the 
code that was used in the EDR and DR1 had scatters of 0.1 mag and
greater, with much more significant offsets. 

The \cmodel\ and PSF magnitudes of stars are 
forced to be identical in the mean by aperture corrections; this was 
true in older versions of the pipeline.  The rms
scatter between model and PSF magnitudes for stars is much reduced, from
0.03 mag to 0.02 mag, the exact values depending on bandpass.
In the EDR and DR1, star-galaxy separation was based on the difference
between model and PSF magnitudes (cf. the discussions in the
EDR paper, Scranton \etal\ 2002, and Strauss \etal\ 2002).  We now do
star-galaxy separation to set the {\tt type} parameter in the pipeline
outputs using the difference between \cmodel\ and PSF
magnitudes, with the threshold at the same value (0.145 
magnitudes).  

Given the excellent agreement between \cmodel\ magnitudes and PSF
magnitudes for point sources, and between \cmodel\ magnitudes and
Petrosian magnitudes (albeit with intrinsic offsets due to aperture corrections)
for galaxies, the \cmodel\ magnitude is now an adequate proxy to use as a
universal magnitude for all types of objects.  As it is approximately
a matched aperture to a galaxy, it has the great advantage over
Petrosian magnitudes, in particular, of having close to optimal noise
properties. 

  For measuring {\em colors} of extended objects, however, we continue
to recommend using
  the model (not the \cmodel) magnitudes; the colors of galaxies were
almost completely unaffected by the DR1 software error (cf., the discussion in
\S~\ref{sec:target}).  The model magnitude is
calculated  using the best-fit
  parameters in the $r$ band, and applies it to all other bands; the
  light is therefore measured consistently through the same aperture
  in all bands.  

\subsection{Other Substantive Changes to the Imaging Pipelines}

\begin{itemize} 

\item The behavior of the deblender of overlapping images has been
further improved since the DR1; these changes are most important for
bright galaxies of large angular extent ($\theta > 1'$).  In the EDR, and to a lesser
extent in the DR1, bright galaxies were occasionally ``shredded'' by the
deblender, i.e., interpreted as two or more objects and taken apart.
With improvements in the code that finds the center of large
galaxies in the presence of superposed stars, and the deblending of
stars superposed on galaxies, this shredding now rarely happens.
Indeed, inspections of
several hundred NGC galaxies shows that the deblend is correct in 95\%
of the cases; most of the exceptions are irregular galaxies of various
sorts.   
\item The PSF is measured from atlas images roughly $7''$ across 
for stars; any error in the sky level determined from these images
couples to spatial variability of the PSF by the
Karhunen-Lo\`eve expansion used to model the PSF.  This manifested
itself in systematic offsets between the PSF and model magnitudes of
stars of several hundredths of a magnitude, even with the fixes in the
model magnitude code described above.  This zero-point term in the PSF
is now explicitly suppressed.

\item The pixel size is $0.396''$, giving well-sampled images for
the typical seeing of $1''$ or more.  On rare occasions when the
seeing became much better than 
$0.9''$ (FWHM), the undersampling causes the code that found stars suitable
for determining the PSF to miss many objects, yielding an incorrect
PSF and therefore poor stellar photometry (the seeing was never good
enough in the runs included in DR1, so this error was not triggered).
Changes to the thresholds for the 
selection of PSF stars have solved this problem. 

\item Astrometry for each object is referred to the reference frame of
  the $r$-band images.  However, for objects of extreme color which
  are undetected in the $r$ band (for example, cool brown dwarfs and
  $z > 5.7$ quasars), DR1 had an error in the astrometric
  transformation from the detection band to the $r$-band, resulting in
  positional errors of several arcseconds.  This problem is fixed in
  DR2, and the positions of objects not detected in the $r$-band are now
  correct.

\item The EDR and DR1 match each SDSS object to the nearest object in
USNO-A2.0 (Monet \etal\ 1998), using a 30 arcsec matching radius.
USNO-A2.0 provides positions at a single epoch (no proper motions are
provided), based on POSS-I plates.  Proper motions are then calculated based
on the SDSS and POSS-I positions, with a typical time baseline of 50 years.
For motions greater than $\sim 40$ mas year$^{-1}$, corresponding to
separations between the SDSS and USNO-A2.0 positions of greater than
2 arcsec, contamination by false matches becomes significant and
rises with increasing motion/separation (see the DR1 web site,
{\tt http://www.sdss.org/dr1}, for a fuller
discussion).  The DR2 reductions use USNO-B1.0 (Monet \etal\ 2003), which
provides positions and proper motions based on various Schmidt
photographic surveys (primarily POSS-I and POSS-II in the area of sky
covered by SDSS).  Each SDSS object is matched to the nearest USNO-B1.0 object
within 1 arcsec, after first converting the USNO-B1.0 positions to the epoch
of the SDSS observations.  This eliminates nearly all of the false matches,
yielding much cleaner samples of high proper motion stars.  The USNO-B1.0
proper motion is then given for each matching SDSS object.  More
sophisticated techniques, using SDSS astrometry to recalibrate the USNO-B1.0
astrometry and then recalculate the proper motions based on both SDSS and
USNO-B1.0 positions, are discussed by Munn \etal\ (2004) and Gould \&
Kollmeier (2004). 


\item When an image is saturated in the SDSS imaging data, the CCD wells
  overflow and a bleed trail results.  However, the total number of
  electrons associated with the object, bleed trail and all, still at
  least approximately reflects the brightness of the object.  For
  objects for which the flag {\tt HAS\_SATUR\_DN} is set in a given
  band, the imaging 
  pipeline includes the counts associated with the bleed trail of
  saturated objects in flux measurements.  In particular, the fiber,
  Petrosian, PSF, and model magnitudes include this light, and it is
  added to the central value of the radial profile (i.e., {\tt
    profMean[0]}).  As the pipeline works on a single frame at a time,
  bleed trails that cross frame boundaries will not be properly
  accounted for. In addition, the fluxes of close pairs of saturated
  stars whose saturated regions overlap will not be correct.

\subsection{Newly Discovered Problems in the Imaging Data}
\label{sec:caveat}

To the best of our knowledge, none of the problems described below is
so severe as to make any substantive change to the conclusions of
science papers using SDSS data.  

\item 
The $u$ filter has a natural red leak
around 7100 \AA\ (cf. Smith \etal\ 2002), which is supposed to be blocked by an interference
coating.  However, under the vacuum in the camera, the wavelength
cutoff of the interference coating has shifted redward (see the
discussion in the EDR paper), allowing some of this red leak through.
The extent of this contamination is different for each camera
column.    It is not completely clear
if the effect is deterministic; there is some evidence that it
is variable from one run to another with very similar conditions in a 
given camera column.  
Roughly speaking, however, this is a 0.02 magnitude effect in the
$u$ magnitudes for 
mid-K stars (and galaxies of similar color),
increasing to 0.06 magnitude for M0 stars ($r-i \approx 0.5$), 0.2
magnitude at $r-i \approx 1.2$, and 0.3 magnitude at $r-i = 1.5$.
There is a large dispersion in the red leak for the redder stars,
caused by three effects: 
\begin{itemize} 
\item The differences in the detailed red leak
response from 
column to column, beating with the complex red spectra of these
objects. 
\item The almost certain time variability of the red leak. 
\item The red-leak images on the $u$ chips are out of focus and
are not centered at the same place as the $u$ image because of lateral
color in the optics and differential refraction--this means that
the fraction of the red-leak flux recovered by the PSF fitting depends
on the amount of centroid displacement. 
\end{itemize}
To make matters even more complicated, this is a {\em detector}
effect.  This means that it is not the real $u-i$ and $u-z$ which drive the
excess, but the instrumental colors (i.e., including the effects of
atmospheric extinction), so the leak is worse at high airmass, when
the true ultraviolet flux is heavily absorbed but the infrared flux
is relatively unaffected.  
Given these complications, we cannot recommend a specific
correction to the $u$-band magnitudes of red stars, and warn the user
of these data about over-interpreting results on colors involving the
$u$ band for stars later than K.

\item There is a slight and only recently recognized downward bias in the
  determination of the sky level in the photometry, at the level of
  roughly 0.1 DN per pixel.  This is apparent if one compares large-aperture
  and PSF photometry of faint stars; the bias is of order 29
  mag arcsec$^{-2}$ in $r$.  This, together with scattered light
  problems in the $u$ band, can cause of order 10\% errors in the
  $u$-band Petrosian fluxes of large galaxies.   

\item The SDSS photometry is intended to be on the AB system (Oke \&
Gunn 1983), by which a
magnitude 0 object should have the same counts as a flat-spectrum source of $F_\nu
= 3631$~Jy.  However, this is known not to be exactly true; the
SDSS photometric zeropoints are slightly off the AB standard. 
We are continuing our effort to pin down these offsets.  Our present estimate,
based on comparison to the STIS standards of Bohlin, Dickinson, \&
Calzetti~(2001) and confirmed by SDSS photometry and spectroscopy of
fainter hot white dwarfs, is that the $u$-band zeropoint is in error
by 0.04 mag, $u_{AB} = u_{SDSS} - 0.04$~mag, and that $g$, $r$, and $i$
are close to AB.  These statements are certainly not precise to better
than 0.01 mag; in addition, they depend critically on the system response of
the SDSS 2.5-meter, which was measured by Doi \etal\ (2004, in preparation).
The $z$-band zeropoint is not as certain at this time, but there is
mild evidence that it may be shifted by about 0.02 mag: 
$z_{AB} = z_{SDSS} + 0.02$~mag.  The large shift in the $u$ band was
expected because the adopted magnitude of the SDSS standard BD+17$^\circ\,4708$ in
Fukugita et al.~(1996) was computed at zero airmass, thereby
making the assumed $u$ response bluer than that at the mean airmass. 
We intend to give a fuller
report on the SDSS zeropoints, with uncertainties, in the near
future.  For further discussion of the conversion between magnitudes
and physical units, see Appendix A. 



\item About 0.3\% of the DR2 imaging footprint area (about 300 out of
  100,000 fields, or 10 deg$^2$) for DR2 are marked as `holes',
  indicated in the CAS by setting quality${}=5$ (HOLE) in the
  `Field' table.  These are areas of sky where no objects are
  cataloged, and researchers interested in structure statistics of
  galaxy or star distributions may wish to mask out these holes from
  their coverage map.  Roughly half of these fields include a very bright
  star (generally $r < 5$) or a very large galaxy or 
  globular cluster, causing the object deblending in the photometric
  pipeline to time out.  While no catalog information is available for
  these fields, the corrected image is available.  Data of
  sufficiently poor quality can also be marked as a hole: very poor
  seeing (significantly worse than $2''$ FWHM), glitches in the telescope
  tracking, and non-photometric data.  There are also a few small
  gaps, also marked as holes,  which fall between two adjacent SDSS
  scans. 

\item The $u$ chip in the third column of the camera is read out on
two amplifiers.  On occasion, electronic problems on this chip 
caused one of the two amplifiers to fail, meaning that half the chip
has no detected objects on it.  This was a problem for only two of the
105 imaging runs included in DR2: run 2190, which includes a total of
360 frames in two separate contiguous pieces on strip 12N (centered
roughly at $\delta = +5^\circ$ in the North Galactic Cap; NGC), and run
2189, which includes 76 frames on stripe 36N near the northern boundary
of the contiguous area in the NGC.  The
relevant frames are flagged as bad in the {\tt quality} flag; in
addition, individual objects in this region have the $u$ band flagged
as {\tt NOTCHECKED\_CENTER} (or, for objects which straddle the
boundary between the two amplifiers, {\tt
  LOCAL\_EDGE}).  Richards \etal\ (2002) describe how the quasar
selection algorithm handles such data; the net effect is that no 
quasars are selected by the $ugri$ branch of the algorithm for these 
data.

\end{itemize}

\section{The SDSS Spectroscopic Data}
\label{sec:spectroscopy}

\subsection{Improvements to Spectrophotometric Calibration}

  There have been three substantial improvements to the algorithms which
photometrically calibrate the spectra (Tremonti \etal\ 2004, in preparation): (1)
improved matching of observed standard stars to models; (2) tying the
spectrophotometry directly to the observed fiber magnitudes from the
photometric pipeline; and (3) no longer using the ``smear''
exposures.  

{\it Analysis of spectroscopic standard stars:}
As described in the EDR paper, each spectroscopic plate contains 16
spectrophotometric standard stars, chosen by their colors to be F
subdwarf stars.  In the EDR and DR1 \texttt{Spectro2d} calibration
pipelines, fluxing was achieved by \emph{assuming} that the mean
spectrum of the stars on each half-plate was equivalent to a synthetic
composite F8 subdwarf spectrum from Pickles (1998).  In the reductions
included in DR2, the spectrum of each standard star is spectrally
typed by comparing with a grid of theoretical spectra generated from
Kurucz model atmospheres (cf. Kurucz 1992) using the spectral 
synthesis code SPECTRUM (Gray \& Corbally 1994; Gray, Graham, \& Hoyt
2001).  The flux calibration vector is derived from the average ratio of
each star (after correcting for Schlegel \etal\ [1998] reddening) and
its best-fit model.

Unlike the EDR and DR1, the 
final calibrated DR2 spectra are \emph{not} corrected for foreground
Galactic reddening (a relatively small effect; the median $E(B-V)$
over the survey is 0.034).  This may be changed in future data
releases. 

{\it Improved Comparison to Fiber Magnitudes:}
We now compute
the absolute calibration by tying the $r$-band fluxes of the standard
star spectra to the fiber magnitudes output by
the latest version of the photometric pipeline.  The latest version
now corrects fiber magnitudes to a constant seeing of $2''$, and
includes the contribution of flux 
from overlapping objects in the fiber aperture; these changes greatly
improve the overall data consistency. 

{\it Smears:}
As the EDR paper
describes, ``smear'' observations are low signal-to-noise ratio (S/N)
spectroscopic exposures made through an effective $5.5'' \times 9''$
aperture, aligned with the parallactic angle.  Smears were designed to
account for object light excluded from the $3''$ fiber due to seeing,
atmospheric refraction and object extent.  However, extensive
experiments comparing photometry and spectrophotometry calibrated with
and without smear observations have shown that the smear correction
provides improvements only for point sources (stars and quasars) with
very high S/N.  For extended sources (galaxies) 
the spectrum obtained in the 3\arcsec\ fiber
aperture is calibrated to have the total flux and spectral shape of
the light in the smear aperture.  This is undesirable, for example,
if the fiber samples the bulge of a galaxy, but the smear aperture
includes much of its disk.  
For this reason, we do {\em not} apply the smear correction to the
data in DR2. 

  To the extent that all point sources are centered in the fibers in
  the same way as are the standards, our flux calibration scheme
  corrects the spectra for losses due to atmospheric refraction without the use of
  smears.  
Extended sources are likely to be slightly over-corrected for
atmospheric refraction.  However, most galaxies are quite centrally
concentrated and more closely resemble point sources than uniform
extended sources.  In the mean, this overcorrection makes the $g-r$
color of the galaxy spectra too red by $\sim1$\%.

\begin{figure}[t]\centering\includegraphics[width=12cm]{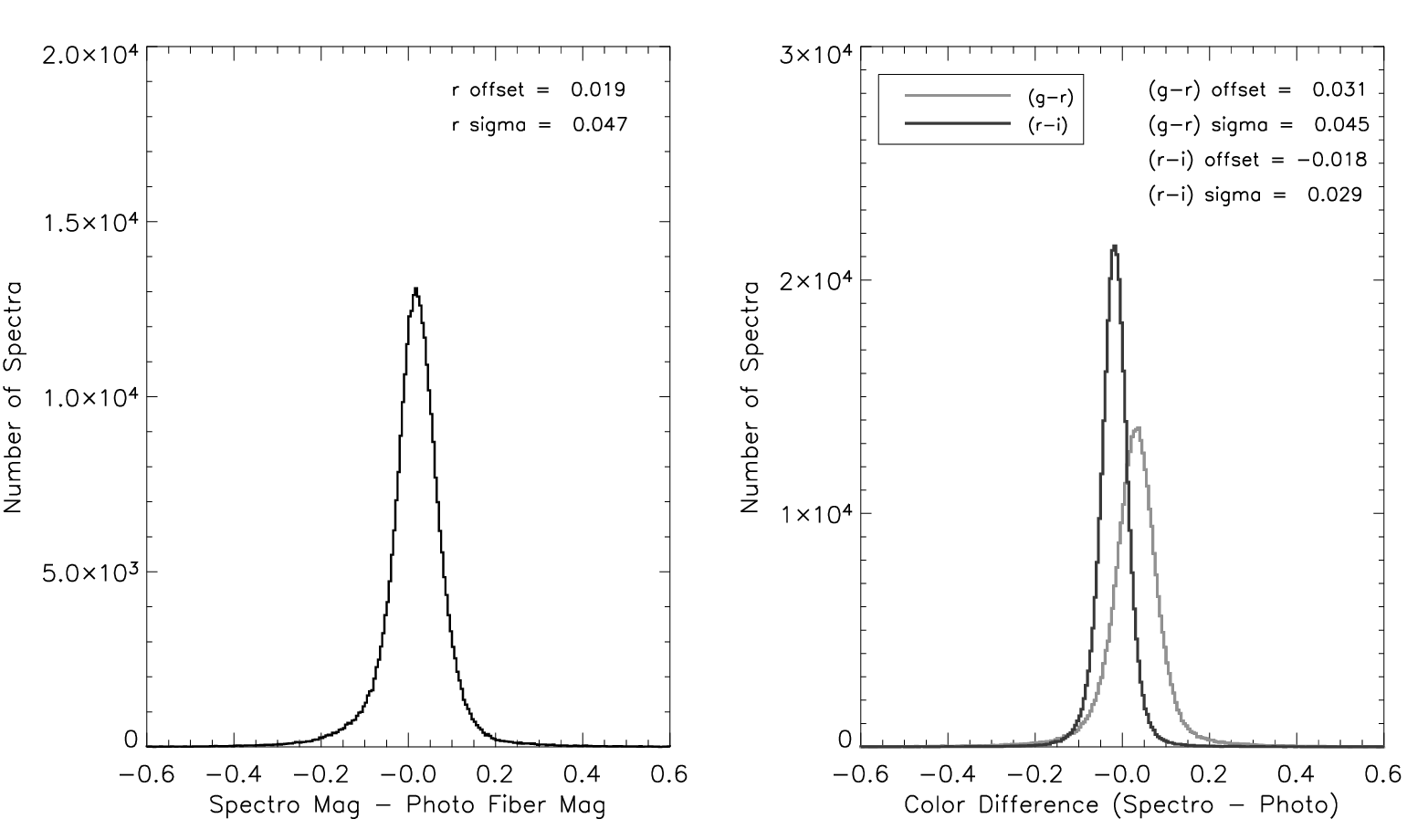}
\caption{Comparison of synthetic $r$ magnitudes and $g-r, r-i$
  colors synthesized from
  the spectra with \texttt{photo} fiber magnitudes.  We have included all
  objects in DR2 with S/N per pixel $>5$.
\label{sphoto1}
}
\end{figure}

The left panel of Figure~\ref{sphoto1} compares $r$ fiber magnitudes with
those synthesized from the spectra of
all DR2 objects with spectral S/N per pixel greater than 5. 
For point sources alone, this
rms difference is 0.040 magnitudes, a 45\% improvement
over DR1.  For
extended sources, the effect of the smears was to give a systematic
offset between spectroscopic and fiber magnitudes of up to a
magnitude; with the DR2 reductions, this trend is gone.  The slight offset of 
the mean from zero is a seeing effect. 

The right panel of
Figure~\ref{sphoto1} compares the $g-r$ and $r-i$ colors derived from
the spectra and photometry; the scatter is $\sim40$\% lower in DR2
than in DR1. The few percent offset of the colors from
zero is an indication that there are small residual errors in our
spectrophotometry, perhaps due to errors in the theoretical models
used to calibrate the standard stars, or to offsets between our
photometric system and a true AB system (see the discussion at the end
of \S~3). 

\begin{figure}[t]
\centering\includegraphics[width=12cm]{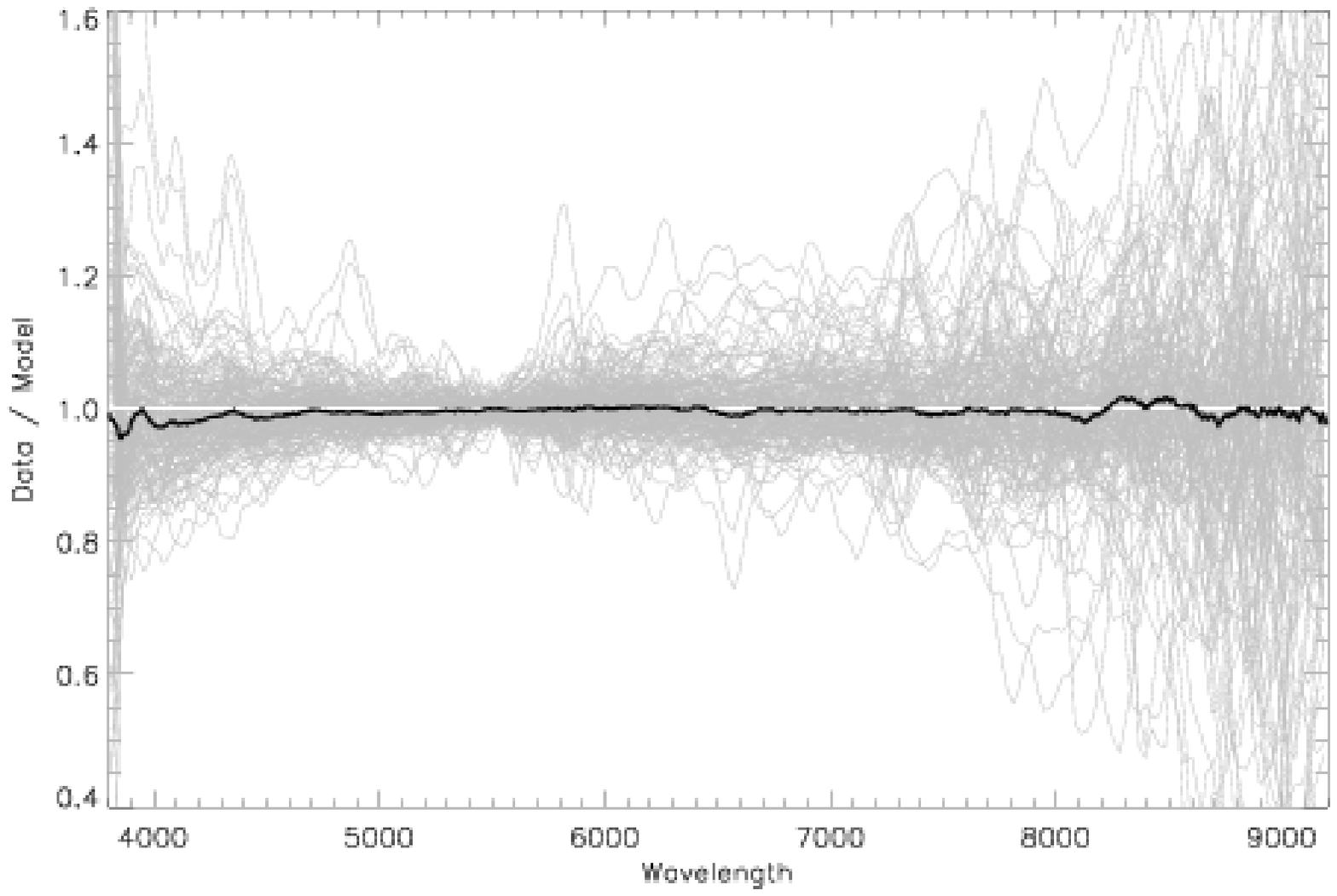}\caption{
Comparison of white dwarf spectra and models.
The grey lines represent 166 individual spectra divided by their best fit
model.  The heavy line is the median.  The equivalent median residuals in DR1
were of order 15\% at 4300\AA; they are now of order a few percent. 
\label{sphoto2}}\end{figure}

To evaluate our spectrophotometry over smaller scales, of order
100~\AA, we compared the calibrated spectra of a sample of 166 hot DA white
dwarfs drawn from Kleinman \etal\ (2004) to theoretical models.  DA white dwarfs
are useful for this comparison because they have simple hydrogen
atmospheres that can be accurately modeled (e.g., 
Finley, Koester, \& Basri 1997). 
Figure~\ref{sphoto2} shows
the results of dividing each spectrum by its best fit model. The
median of the curves shows a net residual of order 2\% at the
bluest wavelengths.  This is a major improvement over DR1 where the
residuals were of order 15\% at 4300\AA\ due to the mismatch between
the observed standard stars and the assumed model.

\subsection{Problematic Spectroscopic Plates}

\begin{table}[hb]
\begin{center}
\caption{{Plates with known problems} \label{badplate}}
\begin{tabular}{ccl}
\hline\hline
Plate ID & MJD & Problem \\
302 & 51668 & Electronic Noise \\
338 & 51694 & Electronic Noise \\
339 & 51692 & Electronic Noise \\
341 & 51690 & Electronic Noise \\
342 & 51691 & Electronic Noise \\
343 & 51692 & Electronic Noise \\
344 & 51693 & Electronic Noise \\
345 & 51690 & Electronic Noise \\
346 & 51693 & Electronic Noise \\
349 & 51699 & Electronic Noise \\
350 & 51691 & Electronic Noise \\
426 & 51882 & Red light leak \\
504 & 52316 & Many rejected pixels \\
721 & 52228 & Many rejected pixels \\
761 & 52266 & Many rejected pixels \\
769 & 52282 & Many rejected pixels \\
770 & 52282 & Many rejected pixels \\
775 & 52295 & Many rejected pixels \\
778 & 52337 & Many rejected pixels \\
\hline
\end{tabular}
\end{center}
\end{table}

A small number of plates, listed in Table~\ref{badplate}, suffered from a
variety of minor problems.
The CCD frames for several plates suffered from a transient electronic
problem in the left amplifier of the red camera in Spectrograph 2
that randomly injected noise into either a single pixel or a cluster of
pixels.  These events mimic cosmic rays and are largely eliminated by
the standard data processing pipeline, but they might leave artifacts in
the reduced spectra.  Another set of plates labeled ``Many Rejected Pixels",
suffered from having the spectrograph collimator improperly focused.
This problem induced a mismatch between the flatfields and the spectra
themselves, causing the optimal extraction process to reject an excessive
number of pixels.  Comparing overlapping objects from adjacent plates
confirms that the redshifts from these problematic plates are
unbiased, but the spectra themselves should not be used for precision 
work or spectrophotometry.  Finally, during the exposure of one plate,
light from an LED somewhere on the telescope found its way to the
spectrographs, resulting in an artificial excess of light centered
roughly at 6500\AA; the spectrophotometry of this plate is quite
poor. 

\subsection{Stellar Radial Velocities}

Spectra for approximately 35,000 Galactic stars of all common spectral
types, targetted both by the quasar target selection algorithm, and in
directed stellar programs, are available with DR2.  Radial velocities
(RVs) are stored as redshifts and were measured by cross-correlation
to a set of stellar templates.
Repeat observations of spectroscopic plates show that the stellar
radial velocities are reproducible to roughly 5 km s$^{-1}$ for stars
brighter than about $r \sim 18$. 
However, the DR2 cross-correlation 
procedure used introduces small additional systematic errors in
addition to possible dispersion errors, depending on spectral type;
these systematics are of order 10 km s$^{-1}$ or less for stars of spectral
type A through K with S/N per resolution element $>10$. 
For white dwarfs and low S/N A stars, lines are too
broad for accurate RV determination, while M dwarf radial velocities
are also less reliable due to prominent molecular bands.  
New for DR2 are measures of the centers and depths of the
Ca triplet lines (8500, 8544, 8664 \AA); these may be useful for more
refined radial velocities of M stars. 

We note that in the EDR and DR1,
zeropoint errors in the templates resulted in quoted RVs for low
metallicity F stars that were too large by 20 km s$^{-1}$, while the
velocities of 
A-type stars were too large by 49 km s$^{-1}$  (Yanny \etal\ 2004).
These problems were uncovered 
through an on-going process involving observations of known RV
standards and cross checks with other model and template fitting
techniques; we hope to continue to improve the tabulated stellar
properties in future data releases or as value added data product
releases. 


\subsection{Mismatches Between the Spectra and Photometric Data}

Each spectroscopic plate includes 32 sky fibers, placed in regions
where the imaging data do not include a detected object; by
definition, there is no photometric object associated with these.
Similarly, there are 437 fibers among the 367,360 spectra in DR2 which
were broken at the time of observation; neither a spectrum nor a
photometric object is associated with them.  Of the remainder, there
are 62 objects for which the matching of fibers to objects cannot be
reconstructed with any confidence, and therefore whose right
ascensions and declinations are uncertain.  For these objects, the
right ascension and declination are listed as $-9999$.

Errors in the deblending algorithm in the {\tt target} reductions caused
spectroscopy to be carried out occasionally on non-existent objects (e.g.,
diffraction spikes of bright stars or satellite trails), or
incorrectly shredded fragments of large galaxies (at $z = 0.01$, a
full 10\% of galaxy targets are such fragments).  Many of these
objects no longer exist in the {\tt best} imaging reductions with its
improvements to the deblender.  In other cases, the photometric pipeline timed
out during the {\tt best} imaging reductions in fields for which {\tt target}
imaging proceeded without problem, so that the {\tt best} photometry is
missing for bona-fide objects. This predominantly happens in fields close to a
few very bright stars.  We expect to recover objects from these ``timeout
holes'' in future data releases.

A total of 663 spectroscopic objects therefore do not have a counterpart in
the {\tt best} images, 0.2\% of the total.  Of these, 80 (including the 62
unmapped fibers) can only be retrieved from the table {\tt specObjAll} in the
database.  The remaining 583 objects are contained in the default
spectroscopic table {\tt specObj}, but will not be found by queries requesting
both photometric and spectroscopic information.

%
%
\subsection{Redshift and Classification Correctness}
As described in the DR1 paper, we have compared the results of two
independent codes to measure the redshifts of the spectra.   These
codes are the pipeline used for the official SDSS reductions, whose
redshifts are based on cross-correlation and emission-line fitting,
and an independent pipeline which uses a
$\chi^2$ method to fit templates directly to spectra (e.g., Glazebrook
\etal\ 1998).  Only 
1.7\% of the objects in DR2 not included in DR1 show gross differences
in redshift and/or classification between the two codes.  We examined
the spectra of all these cases by eye.  One third of those discrepancies
are for very low S/N spectra, for which no redshift is determinable;
in the vast majority of these cases, both pipelines correctly
indicated that they had failed.  Of the remainder, the redshifts or
spectral classifications in the official 
reductions were clearly incorrect in 0.3\% of the spectra, a few
hundred objects in total (many of
them intrinsically interesting objects such as extreme
broad-absorption line quasars, superpositions of objects, and other
oddities).  A list of corrections will be posted to the DR2 web site
when it is completed.  A similar exercise was carried out for the DR1
data (and a similar error rate was found); the resulting corrections
are incorporated into the database. 

\section{Target Selection}
\label{sec:target}
With the change in the model magnitude code (\S~\ref{sec:modelmag}),
the mean $g-r$ and $r-i$ model colors of galaxies have shifted by
about 0.005 magnitudes.  Because the target selection for luminous red
galaxies (LRGs; Eisenstein \etal\ 2001) is very sensitive to color,
this would have increased the 
number density of targets by about 10\%.  Instead, we shifted the LRG
color cuts to compensate; in addition, improved star-galaxy separation
allows tighter cuts on the model-PSF quantity by which stars are
rejected.  Here we give the updated equations 2 and 3
respectively, from Eisenstein \etal\ (2001):
\begin{equation} 
c_\perp  = (r-i) - (g-r)/4.0 - 0.177;
\end{equation}
\begin{equation} 
c_\| = 0.7(g-r) + 1.2[(r-i)-0.177];
\end{equation}
equations 4 and 8 for Cut I: 
\begin{equation} 
r_{\rm Petro} < 13.116 + c_\|/0.3;
\end{equation}
\begin{equation} 
r_{\rm PSF} - r_{\rm model} > 0.24;
\label{eq:star-galaxy}
\end{equation}
and Equations 10, 11, and 13 for Cut II:
\begin{equation} 
c_\perp > 0.449 - (g-r)/6;
\end{equation}
\begin{equation} 
g - r > 1.296 + 0.25(r-i);
\end{equation}
\begin{equation} 
r_{\rm PSF} - r_{\rm model} > 0.4.
\end{equation}
This new version of LRG target selection is applied to the {\tt best}
region of sky reduced with the latest version of the imaging
pipeline.  It is of course not applied retroactively to the {\tt
  target} version of the sky, which used older versions of the
pipeline.  

Due to other subtle differences in the photometric pipeline and the
calibration, these changes will not exactly reproduce the selection criteria
actually used when spectroscopy was carried out.  Indeed, defining an
LRG sample based on the {\tt best} reductions will result in large
spectroscopic incompleteness because so many objects are close to the
boundaries. 
Instead,
one should use the {\tt target} photometry and adjust the calibrations 
of that relative to the {\tt best} calibration.  Of course, if one is
interested in photometric properties of single objects, then we 
recommend the {\tt best} photometry.

The selection of the main galaxy sample (Strauss \etal\ 2002) is based
on Petrosian magnitudes, which have not changed substantially with the
latest versions of the pipelines.  Thus the magnitude limit for this
sample, $r_{\rm Petro} < 17.77$, has not changed.  The improvements to
the model magnitudes have allowed us to tighten the star-galaxy
separation in galaxy target selection; the code uses the same cut as
LRG Cut I (equation~\ref{eq:star-galaxy}).  Note that some of
the EDR and DR1 data were selected with other photometric limits; see
the discussion in Appendix A of Tegmark \etal\ (2004) for details. 

As described in the EDR and DR1 papers and by Schneider \etal\ (2003),
the quasar spectroscopic target selection algorithm has evolved in the
history of the SDSS.  The final version described by Richards \etal\
(2002) went into effect just after the last of the DR1 data were
taken.  Thus all DR2 
data not included in DR1 (i.e., spectroscopic plate numbers greater 
than and including 716) use exactly the algorithm described in
Richards \etal\ (2002).  The most important change implemented at that
time was the addition of sharp color cuts for high-redshift quasars. 

\section{The Future}
    As the name implies, DR2 is the second of a series of releases of what
will eventually be the entire Sloan Digital Sky Survey.  The third data
release, DR3, is planned for late 2004.  DR3 will include all SDSS
survey-quality data taken through June 2003, and will include of order
a 50\% increment over DR2 in spectroscopy and imaging. We expect it to
use essentially the same software used for the processing of DR2. 

\acknowledgements
Funding for the creation and distribution of the SDSS Archive has been
provided by the Alfred P. Sloan Foundation, the Participating Institutions,
the National Aeronautics and Space Administration, the National Science
Foundation, the U.S. Department of Energy, the Japanese Monbukagakusho, and
the Max Planck Society. The SDSS Web site is {\tt http://www.sdss.org/}.

The SDSS is managed by the Astrophysical Research Consortium (ARC) for the
Participating Institutions. The Participating Institutions are The
University of Chicago, Fermilab, the Institute for Advanced Study, the Japan
Participation Group, The Johns Hopkins University, Los Alamos National
Laboratory, the Max-Planck-Institute for Astronomy (MPIA), the
Max-Planck-Institute for Astrophysics (MPA), New Mexico State University,
University of Pittsburgh, Princeton University, the United States Naval
Observatory, and the University of Washington.  We would like to
dedicate this paper to the memory of J. Beverly Oke, whose work on
instrumentation and photometric calibration over the decades was
crucial for the conception and development of the SDSS.

\appendix
\section{Conversion between Magnitudes, Counts, and Fluxes}

As discussed in the EDR paper, the SDSS
catalogs report asinh magnitudes (Lupton, Gunn, \& Szalay 1999)
instead of the conventional logarithmic magnitudes.  The two magnitude
definitions differ only for objects detected at low signal-to-noise
with the SDSS imaging camera.  While asinh magnitudes produce
meaningful colors for such objects, their magnitudes cannot be
converted into Janskys or absolute magnitudes in the same way as those
of brighter objects.  We therefore give detailed instructions here for
the conversion between SDSS asinh magnitudes, Janskys, and imaging
camera counts. For reference, the definition 
of asinh magnitudes is
\begin{equation}
\label{luptitude}
m = -\frac{2.5}{\ln 10} \left[{\rm asinh}\left(\frac{f/f_0}{2\,b}\right) 
+ \ln(b)\right],
\end{equation}
where $f_0$ is the photometric zero point of each filter and $b$ is a
softening parameter as given in Table~\ref{tbl-bprime} (repeated from
the EDR paper). The asinh
magnitude differs by more than 1\% from the conventional magnitude for
objects with flux less than about $10 b f_0$.  For objects brighter
than this, conversion to Jansky and absolute magnitude can be done in
the same way as for conventional magnitudes with less than 1\% error.

\begin{deluxetable}{cccc}
\tablecaption{Asinh Magnitude Softening Parameters \label{tbl-bprime}}
\tablewidth{0pt}
\tablecolumns{4}
\tablecomments{These values of the softening parameter $b$ are set to be
approximate 1$\sigma$ of the sky noise; thus, only low signal-to-noise
ratio measurements are affected by the difference between asinh and Pogson
magnitudes.  The final column gives the asinh magnitude associated
with an object for which $f/f_0=10b$; the difference between Pogson
and asinh magnitudes is $< 1\%$ for objects brighter than this.}
\tablehead{
  \colhead{Band} & \colhead{$b$}  
& \colhead{Zero flux magnitude ($m(f/f_0=0)$)}
	& \colhead{$m(f/f_0=10b)$}
}
\startdata

$u$ & $1.4 \times 10^{-10}$ & 24.63 & 22.12 \cr
$g$ & $0.9 \times 10^{-10}$ & 25.11 & 22.60 \cr
$r$ & $1.2 \times 10^{-10}$ & 24.80 & 22.29 \cr
$i$ & $1.8 \times 10^{-10}$ & 24.36 & 21.85 \cr
$z$ & $7.4 \times 10^{-10}$ & 22.83 & 20.32 \cr

\enddata

\end{deluxetable}



To the extent that the SDSS photometry is on the AB system
(\S~\ref{sec:caveat}), the photometric zeropoint $f_0$ is given by
3631$\,$Jy (Fukugita \etal\ 1996).  With this and the assumption that
the object's spectrum is flat in $f_\nu$, asinh magnitudes can be
converted into Janskies by inverting equation~(\ref{luptitude}).  
A
correction for the spectral shape can be obtained from synthetic
photometry using the SDSS sensitivity curves which are available for
download from the instruments/imager section of the SDSS web site.

To convert between asinh magnitudes and counts $N$ on the SDSS imaging
camera, use Equation~\ref{luptitude}, replacing $f/f_0$ as follows: 
\begin{equation} 
\frac{f}{f_0} \rightarrow \frac{N}{t_{\mathrm{exp}}} 10^{0.4(\mathit{aa} +
  \mathit{kk} \times \mathit{airmass})}
\end{equation}
where $\mathit{aa}$, $\mathit{kk}$, and $\mathit{airmass}$ are the
photometric zeropoint, extinction term and airmass, respectively, for
the object's field and filter from the {\tt field} table or {\tt
tsField} file (these tables also contain the gain value), and
$t_{\mathrm{exp}} = 53.907456$\,s, the exposure time for each SDSS
pixel.


\end{document}